# Challenges in the Deployment of Visuo-Haptic Virtual Environments on the Internet

Jonathan Norman
Computer Science
Armstrong Atlantic State University
Savannah, Georgia
jn2965@students.armstrong.edu

Felix G. Hamza-Lup
Computer Science
Armstrong Atlantic State University
Savannah, Georgia
Felix.Hamza-Lup@armstrong.edu

*Abstract*—Haptic sensory feedback has been shown to complement the visual and auditory senses, improve user performance and provide a greater sense of togetherness in collaborative and interactive virtual environments. However, we are faced with numerous challenges when deploying these systems over the present day Internet. The most significant of these challenges are the network performance limitations of the Wide Area Networks. In this paper, we offer a structured examination of the current challenges in the deployment of haptic-based distributed systems by analyzing the recent advances in the understanding of these challenges and the progress that has been made to overcome them.

*Keywords-haptic feedback; network delay; multimodal; human perception;*

## I. INTRODUCTION

Haptic feedback (simulating the sense of touch) may improve the way humans interact, collaborate, and share information. It offers the ability to augment existing applications and allows for new interactive and collaborative application paradigms not previously possible. The use of haptic feedback, in addition to the visual and auditory modalities, has been shown to enhance operator performance and allow for a greater sense of togetherness among individuals taking part in a collaborative task [1].

Haptic (or force) feedback complements the visual and auditory modalities and increases the safety and performance of Unmanned Aerial Vehicle (UAV) teleoperation [2]. The use of haptic interfaces for interaction with Virtual Environments (VEs) also has numerous applications in several fields including medicine, education, entertainment and Computer-Aided Design (CAD) [3].

Distributed Haptic-based Virtual Environments (DHVEs) are environments allowing an operator to remotely interact with elements in the VE through the use of Haptic Interface Devices (HIDs). The HID allows an operator to physically interact with elements inside the VE and other operators to accomplish a collaborative goal. DHVEs have been successfully deployed over local area networks, employing Quality of Service (QoS) contracts. Large scale experiments involving high-speed test networks have also been successful, though several challenges still exist that prevent the wide-scale deployment of DHVEs over the present day Internet.

We discuss current challenges in the deployment of haptic-based distributed systems from the human perceptual limitation perspective as well as from the network impairments perspective. In Section 2, we examine individual network impairments and discuss how each of them can impact the performance of DHVEs. In Section 3, we review early contributions related to the human perception of the haptic interaction, some of the cognitive limitations concerning haptic interaction, and recent progress in understanding the cross-modal effects of network delay. In Section 4, we review mechanisms that can be employed to conceal and compensate for the impacts of network impairments. In Section 5, we describe an experiment in which we investigate two mechanisms of decreasing the amount of network throughput required by our B.A.C.H. DHVE system [1]. In Section 6, we draw conclusions and comment on the future of DHVE

## II. NETWORK IMPAIRMENTS

### A. Network Performance Metrics

The network performance limitations and unpredictability of large networks, like the Internet, have been the most significant challenges in deploying large scale DHVE applications. Network performance is characterized by a number of metrics such as delay, jitter, loss, and throughput. Impairments across each one of these metrics can impact haptic interactions in different ways [4]. In this section, we introduce and define the standard network performance metrics of delay, jitter, packet loss, and throughput then discuss results from the recent literature concerning their impact on the DHVEs.

### B. The impact of network delay on haptic-enabled systems

Delay is the most common type of network impairment. It is composed of transmission, propagation, queuing and processing delays. Delay has been shown to have a degrading effect on the overall quality of DHVEs, causing an operator to receive sensory feedback prior to, or following their interaction with the VE [5]. Delays can also cause unnatural behavior such as the penetration of solid objects and variations in the object's apparent mass. These inconsistencies can cause the operators to lose their sense of presence in the DHVE thereby compromising its effectiveness.

## C. The impact of network jitter on haptic-enabled systems

Jitter is the statistical variation in network delay. It is measured as the variance of packet interarrival time. Compared to delay, DHVEs have particularly low levels of tolerance to jitter [6]. Numerous results in the recent literature have shown that even under relatively small amounts of jitter, the impact on DHVE applications can be quite severe [4,5,6,7]. To illustrate the relatively low amount of jitter that can be tolerated by haptic systems, Voice over IP (VoIP) and video streaming applications typically require less than 30ms of jitter, while haptic systems become problematic with jitter as low as 2ms.

## D. The impact of network loss on haptic-enabled systems

Loss is a network performance metric which measures the percentage of the number of packets that are lost to the total number of packets sent. In the case where a DHVE application sends VE state updates by transmitting force data over the network, packet loss can reduce the magnitude of force feedback experienced by the operator. For DHVE applications that transmit updates containing object-position data, packet loss can cause an operator to experience abrupt force feedback. When updates provide only relative changes to the state of the remote DHVEs, loss can cause significant divergences in their state [8]. Loss can also cause the apparent weight of objects to vary and can result in the loss of contact between the Haptic Interface Point (HIP) and the objects in the VE [4]. The impact of loss can vary among DHVE implementations. Factors such as the architectural design of the DHVE, whether Client-Server (CS) or Peer-to-Peer (P2P), and the transport protocol it uses, whether TCP or UDP, will each alter the way loss impacts a specific DHVE implementation [8]. It is also worth noting that the occurrence of packet loss is rarely unrelated to delay and jitter. Packet loss is most likely to occur at times of peak network utilization when large delays and jitter are also common.

## E. The impact of network throughput on haptic-enabled systems

Throughput is a measure of the quantity of data that can be transmitted per unit time. Throughput is closely related to delay, jitter and packet loss because when the transmission rate exceeds the available throughput of the communication channel this can lead to delay, jitter and loss. In the case where a DHVE application uses UDP as the transport protocol, packets are dropped when the transmission rate exceeds the available bandwidth or throughput. For DHVE applications that use TCP at the transport layer, when the transmission rate exceeds the amount of throughput available, congestion can occur, causing the TCP algorithm to react by slowing the transmission rate of the sender, resulting in increased delay. In Section 5, we present experimental results of two compensation techniques that have been suggested to lessen the amount of throughput required by networked haptic environments [9]. In these experiments, we evaluate the throughput requirements of a custom cube stacking DHVE application (or task) with and without the compensation techniques in place.

## III. HUMAN PERCEPTUAL LIMITATIONS

### A. The Human Element

The effective realism of a DHVE is very important. An overarching requirement of DHVE design, is to ensure that realism is maintained and that the user does not exhibit disengagement due to unnatural or non-smooth sensory feedback or even imperceptibly fast signals. The way network impairments impact the user's haptic experience is known to vary according to interaction complexity and also movement types. These impacts can also be influenced by cross-modal interactions when haptic channels are combined with the auditory or visual modalities [10]. When considering the impact of network impairments on DHVE applications, the haptic channel is known to be much less tolerant to network impairments than the visual and auditory channels [11].

Quantitative analyses of delayed human perception were done as early as the mid 1900s when the Journal of Experimental Psychology published an article on the psychological effects of delayed visual feedback and how it affected perception [12]. It was established that delayed visual feedback caused a breakdown in what is known as proprioception—the ability to sense the position, location, orientation and movement of the body and its parts.

### B. Cognitive Limitations

Through examination of the physiological, psychophysical and neurological aspects of human interactions, researchers have developed guidelines to ensure that visuo-haptic interactions will maintain this sense of realism. In order for the haptic signals to be perceived as distinct, these guidelines require that they occur at least 5.5ms apart. Additionally, these guidelines suggest that it may not be possible to perceive an individual ordering of signals if they are less than 20ms apart [13].

The impact of network impairments can be altered by cross-modal influences. Recent experimental work studies the influence of delay on the perception of surface stiffness in visual-haptic virtual environments. It was shown that delays in the visual modality alone caused an increase in perceived surface stiffness and that delays in the haptic modality alone caused a decrease in perceived surface stiffness. Simultaneous delays in both the visual and haptic modalities are shown to partly cancel out these changes in perceived surface stiffness [14].

## IV. COMPENSATION TECHNIQUES

### A. Introduction

In this section, we review the mechanisms reported in recent literature as ways to conceal and compensate for network impairments. We present an organized collection of compensation techniques for each type of network impairment, followed by a discussion of their effectiveness and the limitations regarding their use. In addition to presenting a structured examination of these compensation techniques, we also present experimental results of techniques that have been implemented to reduce the throughput requirements of a DHVE application.

### B. Delay concealment techniques

Delay concealment techniques are used to compensate or hide the effects of network delays. These techniques have received a lot of attention in the literature. In [3], the stability of force feedback in the face of network delays is maintained by adding dampening factors at three stages of the haptic interaction. Although these techniques improved system stability, large damping limited the performance of the operator's interaction.

In [15], network delays are compensated for by adjusting the force feedback rendered by the HID. A spring-damper model is used and the force feedback is dynamically adjusted according to the measured end-to-end delay. This delay concealment technique is robust to changing network delays but requires the choice of an accurate stiffness coefficient, which can be difficult.

Another delay concealment technique, presented in [16], utilizes a Delay Synchronization Module (DSM) at the server and Delay Modules (DMs) at each client. The idea is to delay haptic rendering at each client (buffering incoming data) until all clients can do so simultaneously. The authors report improved performance when this delay concealment technique is used in their client-server architecture DHVE.

A somewhat different approach to delay concealment is applied through the use of decorators [17]. Instead of concealing the existence of network delays, decorators provide a visual cue to the operator (indicating the amount of delay that is currently being experienced) so that the operator can adjust their actions appropriately.

### C. Jitter concealment techniques

Jitter concealment has received less attention than delay concealment in the literature. Reference [16] presents a jitter concealment technique through the use of a Prediction Module (PM) to compensate for jitter and packet loss in their CS architecture DHVE. The idea behind the PM is that the client expects to receive an update from the server every millisecond. If no packet is received the PM uses a linear algorithm to predict the new state of the VE and update the client according to this prediction.

Smoothed Synchronous Collaboration Transport Protocol (Smoothed SCTP) [18] is another form of jitter concealment that is heavily based on the Synchronous Collaboration Transport Protocol (SCTP) [19] with the addition of a timestamp and a buffering mechanism for jitter concealment. To produce the jitter concealing effect, a small buffer is used at the receiver which delays incoming packets for a short period of time to ensure that they can be delivered to the application at a constant rate. This jitter concealment technique operates at the expense of increased delay and conceals the effects of jitter as long as the jitter does not exceed the delay imparted by the smoothing buffer.

In [20], jitter concealment is done by using a local-lag algorithm similar to Smoothed SCTP. The local-lag technique is shown to have a dramatic effect in concealing jitter and decreasing DHVE instability caused by jitter.

### D. Packet loss concealment techniques

Packet loss is a commonly occurring network impairment that can have a dramatic impact on the realism experienced in a DHVE. In this section, we discuss related work that attempts to compensate or conceal the effects of loss in the network by employing what we term a packet loss concealment technique.

Forward Error Correction (FEC) is evaluated in [21] to determine how it can be used for loss concealment. FEC compensates for packet loss by introducing redundancy in the communication channel. FEC reduces loss at the application-level by an order of magnitude without introducing additional delays. The drawback of this approach is that it doubles the amount of throughput necessary. Using triple FEC, the loss rate is reduced by an additional order of magnitude while doubling the required throughput again.

The SCTP protocol was developed to provide an architecture that supports tightly coupled collaborative tasks to be performed efficiently in virtual environments [19]. It does so by providing unreliable delivery for normal updates and reliable delivery for important updates. It has been shown to improve collaboration effectiveness in the face of network loss. However, it is susceptible to both jitter and cascading packet loss [22].

Another type of loss concealment, which can also conceal the effects of jitter, is known as a predictor [23]. When DHVE updates are unavailable due to loss or jitter, this can be used to predict the change in state that is most likely to occur. The state of the local VE is then updated according to this prediction. In [22], it is shown that using a linear prediction algorithm to replace lost updates improves collaboration more than using SCTP alone. It is also suggested that using SCTP in addition to a predictor provides the best result, since these methods complement one another.

### E. Throughput concealment techniques

Throughput (or bandwidth) is a measure of the amount of data that can be transmitted over a network per unit time. Throughput is often a defining characteristic of a network communication channel. Insufficient throughput has been shown to introduce higher latencies and packet loss [7]. Because of this, it is not fruitful to discuss throughput concealment techniques as we have done with the other network impairments. Instead of discussing techniques to conceal the limited throughput of a network channel, we present results of an experimental study on techniques for reducing the amount of throughput required by the DHVE.

## V. EXPERIMENTAL ANALYSIS OF TWO THROUGHPUT CONCEALMENT TECHNIQUES

### A. Introduction

We present the experimental results and then analysis of two throughput reduction techniques incorporated into our custom built DHVE in [1]. B.A.C.H. is a CS architecture DHVE in which two participants collaborate to solve a complex cube stacking task (Fig. 1). Each participant controls one haptic pointer, requiring them to coordinate their moves in order to elevate and stack each cube.

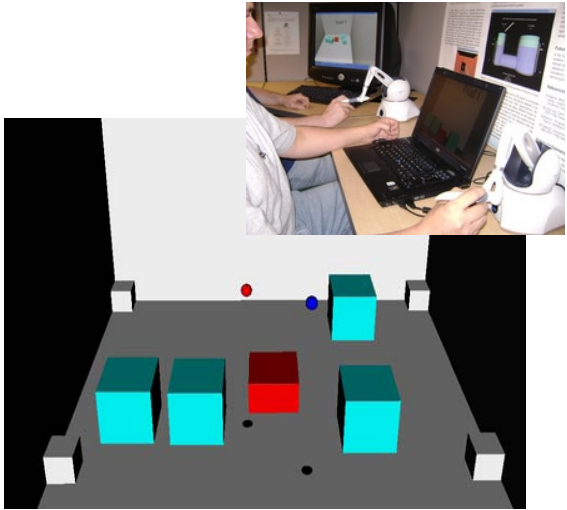

Figure 1. Collaborative Cube Stacking Task

## B. Experiment

During the experiment, two trials were conducted on a Local Area Network (LAN) with two clients and one server. In the first trial, coordinates of the objects in the VE were communicated in update packets over the network with a precision of 1/10000 of a metric unit. In the second trial, the precision was reduced to 1/1000, causing updates to be transmitted less frequently. A third trial was carried out to asses how packet size reduction would impact the throughput requirements of the B.A.C.H. DHVE.

## C. Results

In Table 1, we show the results of the first trial—the throughput under a typical precision setting of 1/10000 of a unit. The transfer rate, given in packets per second, was measured in 10 seconds intervals throughout the length of the experimental trial. The transfer rate is then calculated for each interval and the average of these is given in Table 1 along with the standard deviation and maximum transfer rate across all intervals. Separate measurements, as well as calculations of the required bandwidth (based on an average packet size of 100 bytes), are provided for both inbound and outbound network flows at the server.

TABLE I. BASELINE THROUGHPUT CHARACTERISTICS

| Precision | 1/10000 | | |
|---|---|---|---|
| | *Average* | *Standard Deviation* | *Maximum* |
| Packtets/sec | 384 | 110 | 695 |
| | | | |
| Packets/sec From Server | 133 | 81 | 186 |
| Bandwidth From Server | 106kbps | 65kbps | 149kbps |
| | | | |
| Packets/sec To Server | 251 | 29 | 509 |
| Bandwidth To Server | 201kbps | 23kbps | 407kbps |

Transfer rates and bandwidth calculations with normal precision settings (1/10000 of a unit).

Table 2 shows the decreased transfer rate and bandwidth consumption after reducing the update precision to 1/1000 of a unit.

TABLE II. THROUGHPUT UNDER PRECISION REDUCTION

| Precision | 1/1000 | | |
|---|---|---|---|
| | *Average* | *Standard Deviation* | *Maximum* |
| Packtets/sec | 196 | 19 | 246 |
| | | | |
| Packets/sec From Server | 93 | 10 | 124 |
| Bandwidth From Server | 75kbps | 8kbps | 99kbps |
| | | | |
| Packets/sec To Server | 103 | 12 | 123 |
| Bandwidth To Server | 82kbps | 9kbps | 98kbps |

Transfer rates and bandwidth calculations with reduced precision of 1/1000 of a unit.

In the third experimental trial, we evaluated the impact of reducing the packet size of individual updates. To reduce the size of update packets, the B.A.C.H. application was modified so that the coordinates (transmitted in update packets) contained fewer decimal places than before. In the previous experimental trial, which focused on precision reduction, each coordinate was specified to 12 decimal places. To reduce the packet size for the third experimental trial, the coordinates transmitted in update packets were specified to only four decimal places. This reduced the average packet size from 100 bytes to 73 bytes. Table 3 shows the total reduction in throughput achieved when both throughput reduction techniques are used simultaneously. Bandwidth calculations are based on the average packet size of 73 bytes.

TABLE III. THROUGHPUT UNDER PRECISION REDUCTION AND PACKET SIZE REDUCTION

| Precision | 1/1000 | | |
|---|---|---|---|
| | *Average* | *Standard Deviation* | *Maximum* |
| Packtets/sec | 196 | 19 | 246 |
| | | | |
| Packets/sec From Server | 93 | 10 | 124 |
| Bandwidth From Server | 54kbps | 6kbps | 72kbps |
| | | | |
| Packets/sec To Server | 103 | 12 | 123 |
| Bandwidth To Server | 60kbps | 7kbps | 72kbps |

Transfer rates and bandwidth calculations with reduced precision of 1/1000 of a unit and average packet size reduced to 73 Bytes.

## VI. CONCLUSION

The use of haptic sensory feedback promises to revolutionize the way we interact and collaborate. However, the limited performance characteristics of large networks, such as the Internet, present significant challenges for the research community. To date, the deployment of haptic enabled systems has largely been limited to small networks and high speed dedicated networks, where impairments are less significant. In this paper, we offer a brief structured examination of these challenges and present the recent advances in understanding and addressing these challenges. We identify the individual challenges posed by each type of network impairment, discuss the physiological and psychological limitations of human perception, and identify techniques that can be used to compensate for each type of impairment. We analyze experimental results that illustrate the effects of packet size reduction, as well as update precision reduction, at lowering the amount of throughput necessary for effective haptic collaboration. The use of QoS mechanisms have been shown to provide the higher levels of service needed for haptic collaboration, but are only applicable to networks under a single administrative domain. Further research is necessary to determine whether compensation techniques are sufficient in large networks when QoS guarantees are unavailable.

REFERENCES

[1] B. M. Lambeth, J. LaPlant, E. Clapan, and F. G. Hamza-Lup, "The effects of network delay on task performance in a visual-haptic collaborative environment," Proc. 47th Annual Southeast Regional Conference (ACM-SE 47), ACM, March 2009, pp. 1-5, doi:10.1145/1566445.1566527.

[2] T. M. Lam, M. Mulder, and M. M. van Paassen, "Haptic Feedback in Uninhabited Aerial Vehicle Teleoperation with Time Delay," Journal of Guidance, Control, and Dynamics, vol. 31(6), 2008, pp. 1728-1739, doi:10.2514/1.35340.

[3] J. Kim, H. Kim, B. K. Tay, M. Muniyandi, M. A. Srinivasan, J. Jordan, J. Mortensen, M. Oliveira, and M. Slater, "Transatlantic touch: a study of haptic collaboration over long distance," Presence: Teleoper. Virtual Environ., vol. 13(3), Jul. 2004, pp. 328-337, doi:10.1162/1054746041422370.

[4] A. Marshall, K. M. Yap, and W. Yu, "Providing QoS for Networked Peers in Distributed Haptic Virtual Environments," Advances in Multimedia, vol. 2008, Jun. 2008, Article ID 841590, 14 pages, doi:10.1155/2008/841590.

[5] K. M. Yap, A. Marshall, W. Yu, G. Dodds, Q. Gu, and R. Souayed, "Characterising Distributed Haptic Virtual Environment Network Traffic Flows," Proc. IFIP Network Control and Engineering for QoS, Security and Mobility, Springer Boston, May. 2007, pp.297-310, doi:10.1007/978-0-387-49690-0.

[6] R. T. Souayed, D. Gaiti, W. Yu, G. Dodds, and A. Marshall, "Experimental Study of Haptic Interaction in Distributed Virtual Environments," Proc. EuroHaptics 2004, Springer-Verlag, 2004, pp. 260-266.

[7] K. S. Park and R. V. Kenyon, "Effects of Network Characteristics on Human Performance in a Collaborative Virtual Environment," Proc. IEEE Virtual Reality (VR 99), IEEE Computer Society, Mar. 1999, p. 104, doi:10.1109/VR.1999.756940.

[8] J. Marsh, M. Glencross, S. Pettifer, and R. Hubbold, "A Network Architecture Supporting Consistent Rich Behavior in Collaborative Interactive Applications," IEEE Transactions on Visualization and Computer Graphics, vol. 12(3), May. 2006, pp. 405-416, doi:10.1109/TVCG.2006.40.

[9] M. Fujimoto and Y. Ishibashi, "Packetization interval of haptic media in networked virtual environments," Proc. 4th ACM SIGCOMM Workshop on Network and System Support For Games (NetGames 05), ACM, Oct. 2005, pp. 1-6, doi:10.1145/1103599.1103625.

[10] C. Jay, M. Glencross, and R. Hubbold, "Modeling the effects of delayed haptic and visual feedback in a collaborative virtual environment," ACM Trans. Comput.-Hum. Interact (TOCHI), vol. 14(2), Aug. 2007, Article 8, doi:10.1145/1275511.1275514.

[11] A. Marshall, K. M. Yap, and W. Yu, "Quality of service issues for distributed virtual environments with haptic interfaces," Proc. Multimedia Signal Processing (MMSP 08), IEEE Signal Processing Society, Oct. 2008, pp. 40-45, doi:10.1109/MMSP.2008.4665046.

[12] R. Held, A. Efstathiou, and M. Greene, "Adaptation to displaced and delayed visual feedback from the hand," in Journal of Experimental Psychology, vol. 72(6), Dec 1966, pp. 887-891.

[13] K. S. Hale and K. M. Stanney, "Deriving Haptic Design Guidelines from Human Physiological," IEEE Comput. Graph. Appl., vol. 24(2), Mar. 2004, pp. 33-39. doi:10.1109/MCG.2004.1274059.

[14] B. Knoerlein, M. D. Luca and M. Harders, "Influence of Visual and Haptic Delays on Stiffness Perception in Augmented Reality," Proc. IEEE International Symposium on Mixed and Augmented Reality (ISMAR 2009), Science and Technology Proceedings, Oct. 2009, pp. 49-52, doi:10.1109/ISMAR.2009.5336501.

[15] M. Fujimoto and Y. Ishibashi, "A compensation scheme for network delay jitter of haptic media in networked virtual environments," Proc. 8th World Multi-Conference on Systemics, Cybernetics and Informatics (SCI 2004), vol. 3, 2004, pp. 30-34.

[16] Y. You, M. Y. Sung, and K. Jun, "An integrated haptic data transmission in haptic collaborative virtual environments," Proc. 6th IEEE/ACIS International Conference on Computer and Information Science (ICIS 07), IEEE Computer Society, Jul. 2007, pp. 834–839, doi:10.1109/ICIS.2007.58.

[17] C. Gutwin, S. Benford, J. Dyck, M. Fraser, I. Vaghi, and C. Greenhalgh, "Revealing delay in collaborative environments," Proc. Conference on Human Factors in Computing Systems (SIGCHI 04), ACM, Apr. 2004, pp. 503-510, doi:10.1145/985692.985756.

[18] S. Dodeller and N. D. Georganas, "Transport layer protocols for telehaptics update message," Proc. 22nd Biennial Symposium on Communications, Queen's Univeristy, Canada, 31 May-3 June.

[19] S. Shirmohammadi and N. D. Georganas, "An end-to-end communication architecture for collaborative virtual environments," Comput. Netw., vol. 35(2-3), Feb. 2001, pp. 351-367, doi:10.1016/S1389-1286(00)00186-9.

[20] H. A. Osman, M. Eid, and A. E. Saddik, "Evaluating ALPHAN with Multi-user Collaboration," Proc. 2008 12th IEEE/ACM International Symposium on Distributed Simulation and Real-Time Applications (DS-RT 08), IEEE Computer Society, Oct. 2008, pp. 181-186, doi:10.1109/DS-RT.2008.23.

[21] T. Hudson, M.C. Weigle, K. Jeffay, and R. Taylor II, "Experiments in best-effort multimedia networking for a distributed virtual environment," in Multimedia Computing and Networking, vol. 4312, W.-c. Feng and M. G. Kienzle, Eds. San Jose, CA, Jan. 2001, pp. 88-98.

[22] A. Boukerche, S. Shirmohammadi, and A. Hossain, "Moderating Simulation Lag in Haptic Virtual Environments," Proc. 39th Annual Symposium on Simulation (ANNS 2006), IEEE Computer Society, Apr. 2006, pp. 269-277, doi:10.1109/ANSS.2006.31.

[23] C. Gutwin, J. Dyck, and J. Burkitt, "Using cursor prediction to smooth telepointer jitter," SIGGROUP Bull., vol. 24(1), Apr. 2003, pp. 17-17, doi:10.1145/1027232.1027280.